# Anharmonic stabilization of ferrielectricity in CuInP$_2$Se$_6$

Nikhil Sivadas,[1,*] Peter Doak,[1,2] and P. Ganesh[1,†]

[1]*Center for Nanophase Materials Sciences, Oak Ridge National Laboratory, Oak Ridge, Tennessee 37831, USA*
[2]*Computational Science and Engineering Division, Oak Ridge National Laboratory, Oak Ridge, Tennessee 37831, USA*

Using first-principles calculations and group-theory-based models, we study the stabilization of ferrielectricity (FiE) in CuInP$_2$Se$_6$. We find that the FiE ground state is stabilized by a large anharmonic coupling between the polar mode and a fully symmetric Raman-active mode. Our results open possibilities for controlling the single-step switching barrier for polarization by tuning the Raman-active mode. We discuss the implications of our findings in the context of designing next-generation optoelectronic devices that can overcome the voltage-time dilemma.

DOI: [10.1103/PhysRevResearch.4.013094](10.1103/PhysRevResearch.4.013094)

## I. INTRODUCTION

Recently, ferroelectricity in layered van der Waals materials has attracted a lot of attention because of its applications in high-density nonvolatile memory devices [1,2]. However, very few layered materials demonstrate switchable out-of-plane polarization [3–8]. Of those, the transition metal thiophosphates which includes CuInP$_2$Se$_6$ (CIPSe) are a promising family of materials that host out-of-plane ferrielectricity (FiE) with large values of polarization [4,9,10]. The FiE phases in this family is characterized by a negative piezoelectric response [11,12] and a negative electrostriction [13], along with large dielectric tunability [14]. This led to device applications such as ferroelectric (FE) tunnel junctions [15] and FE field-effect transistors [16,17].

CIPSe was reported to undergo a broad phase transition (200 to 240 K) from the paraelectric (PE) phase (space group $P\bar{3}1c$) to the FiE phase (space group $P31c$) via an incommensurate phase where there is coexistence of FiE and antiferroelectric order [18,19]. Similar to other materials in this family [20], the phase transition is expected to have a large order-disorder character in addition to a displacive character [21]. A coupled oscillator model with an on-site potential that describes the local distortions leading to the double-well potential and an inter-oscillator coupling term that described the ordering of dipoles between the different sublattices can together reconcile with both the displacive character and the order-disorder character in these materials [20,22]. While considerable order-disorder character is expected for the phase transition near the critical temperature, the switching under a finite field from -FiE to +FiE at low temperatures should necessarily be described by only the on-site potential. Here, we assume that the energy barrier for switching corresponds to the energy difference between the FiE phase and the PE phase.

The prior works discussed the stability of the FiE phase for various thicknesses under different boundary conditions [10]. They also obtained the on-site potential using a Landau expansion about the polar off-centering of only the Cu atoms. While the distortions of the Cu sites relative to the PE phase are large (1.2 Å) [23] it would be surprising for such a large distortion to stabilize the FiE phase without the participation from the other atoms. In fact, to understand the microscopic origin of the polarization one needs to include the distortions of the In atoms in addition to the Cu atoms [23]. *As such, a better model for the onsite potential in terms of all the relevant atomic distortions is needed to understand the stabilization of the FiE phase.*

In this paper, we address these questions and discuss the different structural distortions leading to the stabilization of the FiE phase. By choosing an ordered PE parent structure we discuss the structural distortions in the FiE phase using density functional theory (DFT) calculations and group-theoretical methods. We find that the polar displacements of the Cu and In atoms create most of the polarization in the FiE phase. However, this polar distortion alone does not lead to a gain in the energy, and even the full polar mode leads to a very shallow double-well potential (2 meV/f.u.). We report that a strong anharmonic coupling between the polar mode and a fully symmetric Raman-active mode is necessary to stabilize the polar phase. While such a coupling is symmetry allowed even in conventional FEs, we find that its magnitude is large in CIPSe. By analyzing the various contributions to the total energy we discuss the microscopic origin of this large coupling. We demonstrate that strain can be an effective knob to both enhance and suppress the FiE phase. Finally, we discuss the possible implications of this large nonlinear coupling in device applications.

*sivadasn@ornl.gov
†ganeshp@ornl.gov

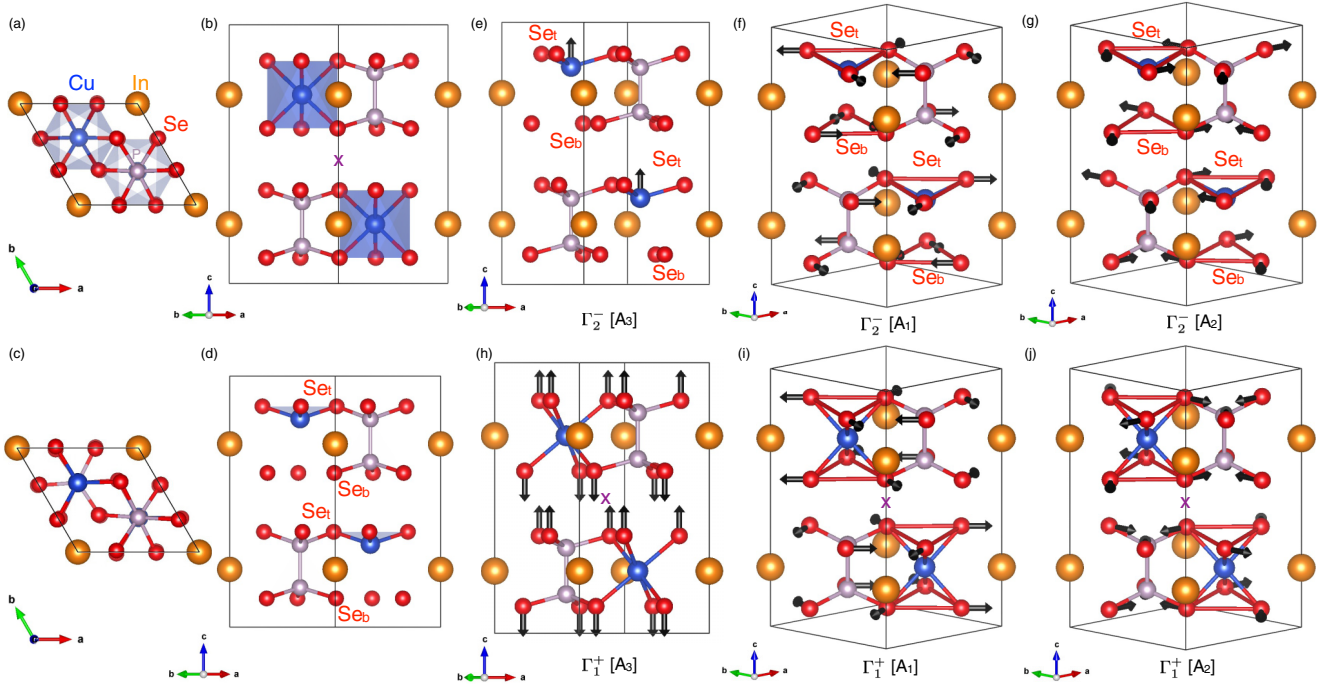

FIG. 1. The crystal structure and the atomic distortions in the paraelectric (PE) and ferrielectric (FiE) phases of CuInP$_2$Se$_6$ (CIPSe). (a) The top view and (b) the side view of the PE phase ($P\bar{3}1c$) with one of the inversion center labeled ("×"). (c) The top view and (d) the side view of the FiE phase ($P31c$). The Se-Se bonds are shown only as visual aid for the modes involving the Se atoms. The FiE phase can be decomposed into six polar symmetry adapted modes (SAMs) and four fully symmetric SAMs that transform as the $\Gamma_2^-$ and $\Gamma_1^+$ irreducible representation (irrep) of $P\bar{3}1c$, respectively. The arrows represent the local distortions. All the arrows have the same amplitude. (e) The displacement of the Cu atom towards the top trigonal Se$_t$ plane ($\Gamma_2^-[A_3]$). (f) The expansion (contraction) of the trigonal Se$_t$ (Se$_b$) planes ($\Gamma_2^-[A_1]$), and (g) the in-phase rotation of the Se$_t$ and Se$_b$ planes about the Cu atoms ($\Gamma_2^-[A_2]$). (h) The out-of-plane expansion of the Se sub-layers ($\Gamma_1^+[A_3]$), (i) an in-plane expansion of the trigonal Se planes ($\Gamma_1^+[A_1]$), and (j) the out-of-phase rotation of the Se$_t$ planes and the Se$_b$ planes ($\Gamma_1^+[A_2]$) about the Cu atoms.[1]

## II. METHODS

We computed the first-principles total energies using Vienna *ab initio* simulation package (VASP) [24], with the PBE functional including van der Waals correction as implemented by Grimme (DFT-D2) [25]. We used projector augmented wave pseudopotentials with valance configurations Cu ($3d^{10}4s^1$), In ($5s^25p^1$), P ($3s^23p^3$), and Se ($4s^24p^4$). Structural relaxation was done with a force convergence tolerance of 0.1 meV/Å using a conjugate-gradient algorithm. The convergence criterion for the electronic self-consistent calculations was set to $10^{-8}$ eV. A regular $8 \times 8 \times 4$ $\Gamma$-centered $k$-point grid was used to sample the Brillouin zone with a plane-wave cutoff energy of 600 eV. The soft-modes were computed using the density-functional-perturbation theory. Polarization was calculated using the Berry-phase approach [26]. While reporting the polarization and the distortion amplitudes the center of mass of the P atoms is used as the origin. This choice does not affect our results. The presented values of the Kohn-Sham eignevalues, the Ewald energy, and the Hartree energy were from the values reported in the VASP output. We used the ISOTROPY software suite to aid with the group-theoretic analysis [27].

## III. RESULTS AND DISCUSSION

### A. Crystal structure

We chose the bulk ordered high-symmetry structure of CIPSe with space group $P\bar{3}1c$ [see Figs. 1(a) and 1(b)] as the reference PE structure for studying the FiE phase [29]. Apart from serving as a zero-point reference for polarization this phase also allows us to study the one-step barrier for the switching of polarization at 0 K. The computed lattice parameters and the occupied Wyckoff positions agree well with the average experimental parameters in the $P\bar{3}1c$ phase (see Table I) [28]. The crystal structure is similar to the transition metal trichalcogenides with a trigonal lattice [30,31]. However, two different atoms (Cu and In) occupy the metal site. This naturally creates an inversion-asymmetry within each layer. The Cu and In ions have a nominal charge of +1 and +3, respectively, resulting in an outer-shell electronic configuration of $d^{10}$ for both the atoms. They both are octahedrally coordinated by six Se atoms. The layers are stacked such that the P$_2$Se$_6$ ligand sits on top of the Cu atom from the other layer, thereby recovering inversion symmetry. This interlayer inversion center is labeled ("×") in Fig. 1.

---

[1]The additional three out-of-plane polar modes ($\Gamma_2^-[A_3]$) involving only one of the In, P, or Se atoms, and the out-of-plane breathing modes involving only the P atoms ($\Gamma_1^+[A_3]$: P) are not shown for the sake of clarify.

TABLE I. The computed structural parameters for the PE phase of CIPSe with space group $P\bar{3}1c$ are compared to the corresponding experimental values [28]. $a$ and $c$ are the lattice constants. The Wyckoff positions of the symmetry inequivalent atoms are also listed with the corresponding experimental values shown in brackets.

| Lattice constants | Computed (DFT-D2 [25]) | | Exp. |
|---|---|---|---|
| $a$ (Å) | 6.37 | | 6.39 |
| $c$ (Å) | 13.20 | | 13.34 |
| Atom | Wyckoff site | $x$ | $y$ | $z$ |
| Cu | 2d | 0.667 (0.667) | 0.333 (0.333) | 0.25 (0.25) |
| In | 2a | 0 (0) | 0 (0) | 0.25 (0.25) |
| P | 4f | 0.333 (0.333) | 0.667 (0.667) | 0.164 (0.166) |
| Se | 12i | 0.349 (0.331) | 0.340 (0.340) | 0.120 (0.120) |

The PE phase is unstable with both zone-center and zone-boundary instabilities. As we are interested in understanding the FiE phase which is reported to have the same cell size as the PE phase [11], we only consider the zone-center instability here. We verified that the relaxed structures from the other instabilities have higher energy than the FiE phase.

The relaxed crystal structure of the FiE phase with space group $P31c$ is shown in Figs. 1(c) and 1(d). The computed lattice parameters and Wyckoff positions compare well with the corresponding experimental values (see Table II) [11]. The Cu and In atoms exhibit an antiparallel out-of-plane distortion relative to each other. Further, the P and Se Wyckoff sites are split into two sites each, corresponding to the top and bottom sub-layers within each monolayer. We label the top (bottom) trigonal Se plane within each CIPSe layer as $Se_t$ ($Se_b$).

### B. Mode decomposition of the FiE phase

The polar mode that transforms as the $\Gamma_2^-$ irreducible representation (irrep) of $P\bar{3}1c$ is the primary order parameter for the transition from the PE phase to the FiE phase. There are six symmetry-adapted modes that transform as the $\Gamma_2^-$ irrep. Four of them are labeled $\Gamma_2^-[A_3]$, and involve the displacement of one of the four atom types along the out-of-plane direction. Figure 1(e) shows the $\Gamma_2^-[A_3]$ mode for the Cu atom. In addition, there are two symmetry adapted modes (SAMs) comprising of the in-plane distortions of the Se octahedron. The first mode labeled $\Gamma_2^-[A_1]$ is a breathing mode corresponding to the expansion (compression) of the trigonal $Se_t$ ($Se_b$) plane along (away from) the direction of the Cu displacement [see Fig. 1(f)]. This results in an in-phase rotation of the trigonal Se planes about the $P_2$ bond. The second mode labeled $\Gamma_2^-[A_2]$ is a breathing mode about the $P_2$ bond [see Fig. 1(g)], resulting in an in-phase rotation about the Cu atoms. The polar mode is made up of a combination of these six SAMs.

TABLE II. The computed structural parameters for the FiE phase of CIPSe with space group $P31c$ are compared to the corresponding experimental values [11]. $a$ and $c$ are the lattice constants. The Wyckoff positions of the symmetry inequivalent atoms are also listed with the corresponding experimental values shown in brackets.

| Lattice constants | Computed (DFT-D2 [25]) | | Exp. |
|---|---|---|---|
| $a$ (Å) | 6.43 | | 6.40 |
| $c$ (Å) | 13.27 | | 13.32 |
| Atom | Wyckoff site | $x$ | $y$ | $z$ |
| Cu | 2b | 0.667 (0.667) | 0.333 (0.333) | 0.154 (0.147) |
| In | 2a | 0 (0) | (0) | 0.255 (0.259) |
| $P_t$ | 2b | 0.333 (0.333) | 0.667 (0.667) | 0.327 (0.326) |
| $P_b$ | 2b | 0.333 (0.333) | 0.667 (0.667) | 0.157 (0.158) |
| $Se_t$ | 6c | 0.310 (0.301) | 0.325 (0.323) | 0.105 (0.110) |
| $Se_b$ | 6c | 0.358 (0.355) | 0.008 (0.005) | 0.371 (0.371) |

TABLE III. The amplitude of the various SAMs that transform like the $\Gamma_1^+$ (fully symmetric mode) and the $\Gamma_2^-$ (polar mode) irreps of the PE phase (space group $P\bar{3}1c$) of CIPSe. Figure 1 shows the corresponding distortions.

| SAM in $P\bar{3}1c$ | Amplitude (Å) |
|---|---|
| $\Gamma_1^+[A_3]$: P | −0.01 |
| $\Gamma_1^+[A_3]$: Se | 0.03 |
| $\Gamma_1^+[A_1]$: Se | 0.08 |
| $\Gamma_1^+[A_2]$: Se | 0.03 |
| $\Gamma_2^-[A_3]$: Cu | 1.15 |
| $\Gamma_2^-[A_3]$: In | −0.18 |
| $\Gamma_2^-[A_3]$: Se | 0.05 |
| $\Gamma_2^-[A_1]$: Se | 0.13 |
| $\Gamma_2^-[A_2]$: Se | 0.00 |

In addition, the polar mode induces a Raman-active mode which preserves the full symmetry of the parent space group (labeled $\Gamma_1^+$). This mode in turn is made up of a combination of four SAMs that involve the displacement of only the P and Se atoms. Two of these SAMs are labeled $\Gamma_1^+[A_3]$ and correspond to the out-of-plane expansion of the individual P and Se layers. Figure 1(h) shows this $\Gamma_1^+[A_3]$ mode for the Se atoms. The other two modes involve the in-plane distortion of the trigonal Se planes and are labeled $\Gamma_1^+[A_1]$ and $\Gamma_1^+[A_2]$. The $\Gamma_1^+[A_1]$ mode corresponds to the expansion of both the top and bottom trigonal Se planes about the Cu atoms [see Fig. 1(i)], whereas the $\Gamma_1^+[A_2]$ mode involve the out-of-phase rotation of the top and bottom trigonal Se planes about the Cu atoms [see Fig. 1(j)].

### C. The microscopic origin of polarization

Table III shows the amplitude of the corresponding SAMs in the FiE phase. Notably, the largest distortion involves the out-of-plane motion of the Cu atoms towards the $Se_t$ trigonal face (labeled $\Gamma_2^-[A_3]$: Cu), and has an amplitude of 1.15 Å. Such a large distortion is atypical even for FEs. It is reasonable to assume that this will lead to large anharmonicity in the phonon modes, as we demonstrate later.

Table IV lists the atom-projected polarization with respect to the PE phase. The contributions from both $\Gamma_1^+$ and $\Gamma_2^-$ displacive modes are included. The polar distortions of the Cu and In atoms carry a polarization of 6.17 $\mu$C/cm$^2$ and −2.70 $\mu$C/cm$^2$, respectively. As the Born-effective charge for

TABLE IV. The contribution to polarization from the distortions of the various atoms with respect to the PE phase. The contributions from both $\Gamma_1^+$ and $\Gamma_2^-$ displacive modes are included.

| Atom | Polarization ($\mu$C/cm$^2$) |
|---|---|
| Cu | 6.17 |
| In | −2.71 |
| Se | −0.96 |
| CuIn | 3.29 |
| CuSe | 5.60 |
| InSe | −3.82 |
| CuInSe (FiE-phase) | 2.58 |

In (2.24) is larger than that of Cu (0.90), a relatively small displacement of In atoms (0.18 Å) can carry a large dipole moment. This results in the reported FiE phase with partial cancellation of polarization between the two sites, and a net polarization along the direction of the Cu displacement. The fully relaxed FiE phase has a total polarization of 2.58 $\mu$C/cm$^2$, which compares well with prior first-principles predictions [10]. Even though the polarization is smaller than what is typically observed in conventional FEs, it is larger than other reported ferrielectrics such as ammonium sulphate (0.62 $\mu$C/cm$^2$) [32].

Figure 2 shows the total polarization as a function of the fractional amplitude of the fully symmetric mode ($Q_{\Gamma_1^+}$) and the polar mode ($Q_{\Gamma_2^-}$) in the FiE phase with respect to the PE phase. These modes induce local distortions proportional to that listed in Table III. The lattice parameters are fixed to that of the fully relaxed PE lattice parameters. This does not affect our central results. It is evident that the $Q_{\Gamma_1^+}$ on its own does not carry any polarization, as expected from symmetry. But a condensation of $Q_{\Gamma_1^+}$ on top of the polar mode $Q_{\Gamma_2^-}$ modifies the electronic polarization. Still, the polar mode carries most of the polarization in the system. As the metal atom distortions dominate the polar mode, we can conclude that the microscopic origin of the polarization in FiE CIPSe is from the distortion of the metal atoms. This is also corroborated by Table IV which shows that the distortions of the Cu and In atoms create most of the polarization in FiE CIPSe.

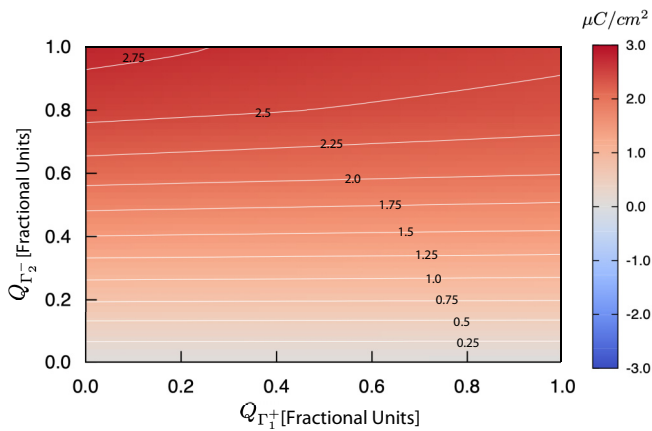

FIG. 2. The decomposition of the total polarization as a function of the fractional amplitude of the fully symmetric mode ($Q_{\Gamma_1^+}$) and the polar mode ($Q_{\Gamma_2^-}$). The polar mode creates most of the polarization. The fully symmetric mode on its own does not carry any dipole moment.

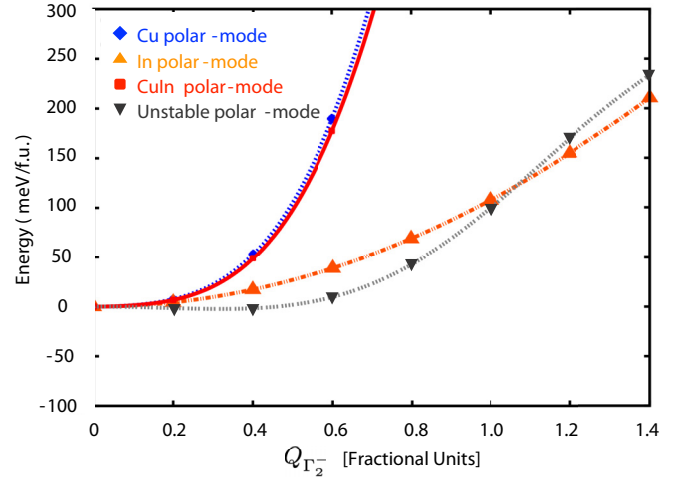

FIG. 3. The energy as a function of the atom-projected polar mode ($Q_{\Gamma_2^-}$). The energy as a function of the full polar mode that is unstable (black), and the component of the mode involving the displacement of only the Cu atoms (blue), the In atoms (orange), and both Cu and In atoms (red). The energy gain from only the polar modes is small.

### D. Energy profile of the polar mode

While the Cu and In atoms lead to most of the polarization in the FiE phase, from Fig. 3 we find that neither the individual distortion of the Cu atoms (blue lines) or In atoms (orange line), nor the collective ferrielectric distortion of both the atoms (red lines) creates a double-well potential with respect to the PE phase [33]. This highlights the inadequacy of models based on a Landau expansion of only the Cu polar off-centering to explain the FiE phase [10]. The energy profile for the full polar-instability in the PE phase (black curve in Fig. 3) shows a shallow double-well potential (2 meV). Here, while we present the total energy as a function of the polar mode along the relaxed PE-FiE direction we verified that similar results are obtained if we choose instead the polar soft-mode eigenvectors from the force-constant matrix (see Fig. S1 in the Supplemental Material [34]). This highlights the importance of the collective distortion of all atoms that transforms as the $\Gamma_2^-$ irrep of $P\bar{3}1c$ rather than just the displacement of Cu (or In) atom as discussed in prior research [10,18,19], as the primary order parameter for understanding the symmetry lowering phase transition in CIPSe.

Interestingly, both the well-depth and amplitude of the polar mode at the minimum are much smaller than that of the fully relaxed FiE phase. The fully relaxed FiE phase has an energy gain of 98 meV/f.u. with respect to the PE phase. This is in sharp contrast to other proper FEs where the dominant polar distortion on its own accounts for a large portion of the energy gain in the polar phase characterized by a deep double-well potential [35]. This large energy difference in CIPSe immediately suggests that to understand the microscopic mechanism that stabilizes the polar phase requires not only the polar mode but also the nonpolar modes.

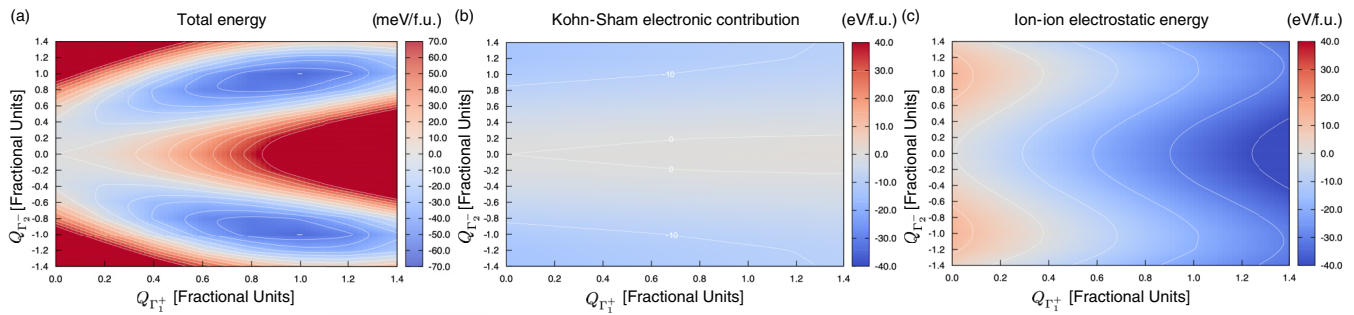

FIG. 4. The anharmonic energy surface. (a) The total energy (meV/f.u. atom), (b) the electronic-structure contribution from the valance electrons (eV/f.u. atom), and (c) the contribution from the ion-ion electrostatic interaction (eV/f.u. atom), as a function of the fractional amplitude of the fully symmetric mode ($Q_{\Gamma_1^+}$) and the polar mode ($Q_{\Gamma_2^-}$). The scale of the latter is the same as that in Fig. 3. All the reported energies are relative to the fully relaxed PE phase ($P\bar{3}1c$).

### E. Total energy surface

To explain the above mentioned discrepancy we compute the energy surface [see Fig. 4(a)] by varying fractional amplitudes of the fully symmetric mode ($Q_{\Gamma_1^+}$) and the polar mode ($Q_{\Gamma_2^-}$). We find that the total energy goes up with $Q_{\Gamma_1^+}$. This is expected as the initial configuration is completely relaxed. We also recover the shallow double-well as we condense $Q_{\Gamma_2^-}$ mode. The lowest energy structures corresponding to the fractional coordinates of $(1, \pm 1)$ have 61 meV/f.u. lower in energy than the PE phase. While further stabilization (37 meV/f.u.) of the FiE phase is provided by a full structural relaxation, Fig. 4(a) shows that the dominant factor leading to the stabilization of the FiE phase is the strong anharmonic coupling between the polar mode and the fully symmetric mode. Such anharmonicity has also been reported in other members of the thiophosphate family [36].

To quantify this anharmonic coupling, we fit the energy surface about the PE phase. The total energy ($E$) can be written as

$$E = E_0 + b_{02}Q_{\Gamma_2^-}^2 + b_{04}Q_{\Gamma_2^-}^4 + b_{06}Q_{\Gamma_2^-}^6 \\ + b_{08}Q_{\Gamma_2^-}^8 + c_{12}Q_{\Gamma_1^+}Q_{\Gamma_2^-}^2 + c_{14}Q_{\Gamma_1^+}Q_{\Gamma_2^-}^4 + a_{20}Q_{\Gamma_1^+}^2, \quad (1)$$

where $E_0$ is the energy of the PE phase that is set to zero, $a_{ij}$ and $b_{ij}$ are the real-valued coefficients of the expansion about $Q_{\Gamma_1^+}$ and $Q_{\Gamma_2^-}$, respectively. The free-energy expansion considered in Eq. (1) is only for the on-site interactions. Additional intersite coupling coming from the zone-boundary modes needed to considered to discuss order-disorder transition [20]. As $Q'$s are in fractional units, the coefficients in the fit have units of energy allowing us to compare the contribution from each term. $c_{ij}'$s are the real-valued coefficients corresponding to the anharmonic coupling between the two modes. Anharmonicity due to coupling between multiple modes has attracted a lot of attention in the context of nonproper FEs [37–41] and antiferroelectrics [42,43]. However, in the case of proper FEs even though such a coupling between the polar mode and the fully symmetric mode is allowed by symmetry they are considered to be small and are typically ignored [44].

Our approach of including this coupling between the full polar mode and other structural distortions goes beyond the existing models which looks at only the Cu distortions [10]. We will show that such an anharmonic coupling is crucial to the stabilization of the polar phase, and is key to controlling the polarization in CIPSe. Table V shows the coefficients obtained from fitting the energy surface in Fig. 4(a). For the total energy to be bounded the coefficients of the largest-order polynomial for each mode should be positive. As the sixth-order coefficient of the polar mode ($b_{06}$) was negative we included up to the eighth-order term. The asymptotic standard error in the fit was less than 8.2%, with most coefficients within 4.5%. The fit can be further improved by adding additional terms (see Table S1 in the Supplemental Material) [34]. As the overall conclusions are independent of these additional terms, we omit them in the discussion of the minimal model.

We find $b_{02}$ to be negative and $b_{04}$ to be positive as expected of a characteristic double-well potential due to a polar-instability [45]. However, from Table V it is clear that the anharmonic coupling is anomalously large with $c_{12}$ about eight times larger than the harmonic term $b_{02}$. For perspective, this is comparable to the improper FE coupling between the polar mode and the zone-boundary mode in YMnO$_3$ [38]. The effective quadratic coefficient ($B_{02}^{\text{eff}}$), including the anharmonic terms from Eq. (1), is

$$B_{02}^{\text{eff}} = \left(b_{02} + c_{12}Q_{\Gamma_1^+}\right). \quad (2)$$

The strong anharmonic coupling renormalizes $B_{02}^{\text{eff}}$ to become more negative. Similar renormalization of the optimal polarization and the Curie-Weiss temperature in conventional oxide FEs was discussed due to electrostriction [45]. But an anomalously large contribution of the fully symmetric mode leading to this renormalization as we find in CIPSe is new.

TABLE V. The value of the coefficients (in meV) in Eq. (1) from fitting the energy surface shown in Fig. 4(a).

| $b_{02}$: −47.1 | $b_{04}$: 245.7 | $b_{06}$: −116.0 | $b_{08}$: 17.3 |
|---|---|---|---|
| $c_{12}$: −356.7 | $c_{14}$: 100.1 | $a_{20}$: 102.6 | |

### F. The microscopic mechanism of the coupling

Prior works discussed the possibility that a filled $d^{10}$ outer-shell electronic configuration is second-order Jahn-Teller active, leading to the off-centering of both the Cu and In atoms in this class of materials [9,46–48]. So to elucidate the physical origin of the large nonlinear coupling between the polar mode and the fully symmetric mode, we analyzed the various contributions to the total energy. Specifically, we focus on the Kohn-Sham eigenvalues and the Ewald energy— i.e., the electrostatic ion-ion interaction terms for the FiE phase relative to the PE phase, as shown in Figs. 4(b) and 4(c), respectively. The former corresponds to the electronic component of the second-order Jahn-Teller effect, and should lead to the lowering of the Kohn-Sham energy. The latter can be compared to the effect of distortion without an electronic redistribution which should lead to an increase in energy. Together they model a second-order Jahn-Teller effect [45]. This is shown as a function of $Q_{\Gamma_1^+}$ and $Q_{\Gamma_2^-}$ in Fig. 4.

We find that the polar mode lowers the total energy by reducing the Kohn-Sham electronic contribution to the energy [see Fig. 4(b)]. But the condensation of the polar mode results in a large Ewald energy penalty [see Fig. 4(c)]. Physically, as the Cu ions move closer to the Se ions [see Fig. 1(e)], thereby lowering the Kohn-Sham energy, the Ewald part of the total energy goes up due to an increased ion-ion repulsion. However, this shorter Cu-Se$_t$ bond can be stabilized by an expansion (contraction) of the trigonal Se$_t$ (Se$_b$) plane. As shown in Table III this distortion of the trigonal Se planes is dominated by the $A_1$ modes of both $\Gamma_2^-$ [see Fig. 1(f)] and $\Gamma_1^+$ [see Fig. 1(i)]. As the expansion of the Se$_t$ plane does not have the same amplitude as the contraction of the Se$_b$ plane as shown in Fig. 1(f) for the $\Gamma_2^-$ $A_1$ mode, in addition to this mode the $\Gamma_1^+$ $A_1$ mode [see Fig. 1(i)] is needed to stabilize the displaced Cu atoms. This naturally leading to a coupling between the two modes. The expansion and contraction of the Se sublayers we find is consistent with prior x-ray diffraction studies in this class of compounds [9]. Further, Fig. 4(c) shows that this coupling between the polar mode and the fully symmetric mode overcomes this energy penalty. So, the stabilization of the FiE phase in CIPSe is mediated by the coupling between the metal distortions and the chalcogen atoms that also contributes to the fully symmetric Raman active mode. Such a competition between the electronic degrees of freedom and the repulsive forces between the ions is in line with a second-order Jahn-Teller effect [9,45–48].

Our analysis gives a clear microscopic mechanism of the unusual coupling between the polar mode and the fully symmetric mode that leads to the stabilization of the FiE phase. We find that this anharmonic stabilization persists even in the monolayer limit (see Fig. S5 in the Supplemental Material [34]).

### G. Strain-control of FiE

We also explored the effect of other distortions that transform as the $\Gamma_1^+$ irrep of the space group $P\bar{3}1c$. Previous works on oxides showed that polar-distortions naturally couple to strain [35,49]. So, we discuss the case of an in-plane biaxial strain ($\epsilon_{x^2+y^2}$), which uniformly expands the 2D-area for positive values. The effect of this strain is similar to that of the displacive $\Gamma_1^+$ mode [see Fig. 1(i)]. While only the Se atoms participate in the displacive mode, the strain mode includes all the atoms. Naturally, we expect this component of the strain to have a larger effect on the anharmonic coupling. We also check the contribution from other fully symmetric strain mode and found that the dominant effect comes from this biaxial strain mode.

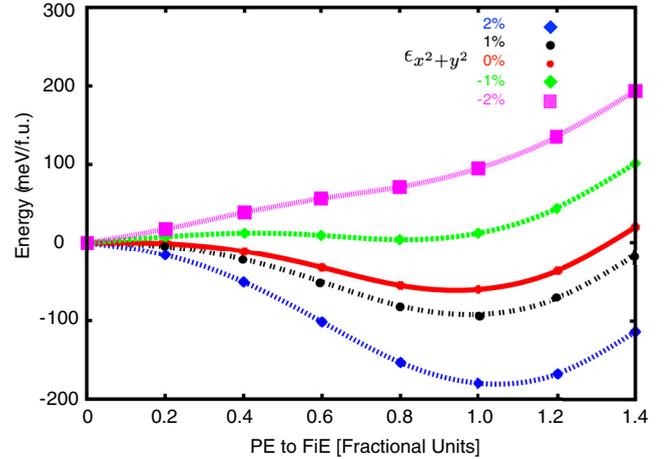

FIG. 5. The total energy as a function of the fractional amplitude of the distortion from the PE phase to the FiE phase, for different value of biaxial strain ($\epsilon_{x^2+y^2}$). Results for tensile strains of 2% (blue), 1% (black), the unstrained case (red), and compressive strain of 1% (green), and 2% (magenta) are reported. The energy difference can be enhanced or suppressed by the application of an in-plane strain.

Figure 5 shows the total energy profile for the FiE phase relative to the PE phase for various value of $\epsilon_{x^2+y^2}$. For the unstrained case (red solid line) we recover the FiE phase as the low-energy state. We find that for a compressive strain of 1% (green line) the FiE phase, although a local minimum, has higher energy (4 meV/f.u.) than the PE phase with a small energy barrier (12 meV/f.u.) between the states. For intermediate values of compressive strain where the FiE remains the ground state but where the energy barrier is comparable to the quantum fluctuations a quantum PE state could be stabilized by strain where quantum fluctuations suppress the ordering of dipole moments [50,51]. Incidentally, a quantum paraelectric phase was proposed in this family of materials driven by chemical substitution [52].

For a compressive strain of 2% (magenta line) the polar instability is fully suppressed and the PE phase becomes the lowest energy structure. On the other hand, for a tensile strain (1% shown in black lines and 2% shown in blue line) the well-depth around the FiE phase becomes deeper. Note that the FiE phase has a $\epsilon_{x^2+y^2}$ strain of $\sim 1\%$ relative to the PE phase (see Tables I and II), and largely accounts for the additional 37 meV gain from relaxing the lattice constants in the FiE phase. As biaxial strain can be easily controlled experimentally instead of directly tuning the displacive $\Gamma_1^+$ mode, mechanical strain can be an effective tool to both suppress and enhance the FiE phase, providing a way to tune the energy barrier corresponding to a single-step switching of the polarization.

## IV. CONCLUSION AND OUTLOOK

While our analysis is done at 0 K in the absence of an external electric field and domains, it gives some preliminary insights into the factors that govern the switching of polarization under an applied electric field. Within a single-step switching process we assume that the energy barrier for switching is determined by the energy difference between the FiE phase and the PE phase. Our analysis, shows that this barrier is determined primarily by a large nonlinear coupling between the polar mode and the fully symmetric Raman-active mode. This opens up possibilities to both enhance and suppress the energy barrier for polarization switching by modifying the fully symmetric mode, instead of the polar mode. Since the fully symmetric mode is Raman active, in addition to mechanical strain, it should also couple to ultra-fast optical probes allowing for the optical control of the switching barrier. [53–55]. The application of such an optical field should be faster than applying a mechanical strain.

When the barrier energy is reduced sufficiently, a faster electrical switching of the polar state can be achieved within the transition state theory, assuming a one-step switching process. On removal of the two fields the higher barrier around the FiE phase, for example in the unstrained case, can be recovered. This will lead to a long-lifetime of the polar phase, which can be even enhanced for instance by engineering a tensile strain as shown in Fig. 5. This understanding can provide a first step to addressing the voltage-time dilemma [56].

We note that this approach for overcoming the voltage-time dilemma is not unique to CIPSe. Other materials where the structural stabilization of the ordered phase involves significant contributions from modes other than the modes describing the memory-order parameter has the potential to overcome this inherent bottleneck. In the case of CIPSe, while the metal polar-distortions created the memory order-parameter the stabilization of the polar phase was due to the anharmonic coupling between the polar mode and the fully symmetric mode. Improper FEs, pseudoproper FEs, hybrid improper FEs and triggered phases are all potential candidates for this as the polarization itself is not the primary order parameter and the stabilization of the polar phase comes from the coupling between polarization and other structural distortions. We note that a strain-induced suppression of the barrier was recently reported in hybrid improper FE [57].

Finally, we comment on the nature of the FiE phase. Within a displacive picture CIPSe is a proper FiE as the polar mode is the primary order parameter and fully determines the symmetries of the FiE phase. However, unlike other proper FEs where the energy lowering term is $P^2$ [35,58–60], the driving term that stabilizes the polar phase is dominated by the anharmonic coupling between the polar mode and the fully symmetric Raman active mode. This distinguishes the ferrielectricity in CIPSe and other layered thiophosphates from other known FEs.

While our discussion is centered around a on-site model for the total energy, our results should also be important in a multi-step switching process. Future works need to include the effect of zone-boundary instabilities also to understand the order-disorder nature of the phase transition and the intermediate incommensurate phase reported at the phase boundary. Similarly, the impact of our findings in the context of the unconventional electromechanical response found in the domain walls in this family of materials, and in the actual path of polarization switching under an applied field also needs to be explored.

In summary, we discuss the microscopic origin of polarization and the structural-stabilization of ferrielectricity in CIPSe. The distortions of the Cu and In atoms create most of the polarization in FiE CIPSe. On the other hand, the anharmonic coupling between the polar mode and the fully symmetric distortions involving the Se atoms stabilizes the FiE phase. The energy barrier for the single-step switching of the polarization in CIPSe can be controlled by changing the fully symmetric mode for instance, by the application of a uniform in-plane strain. This can help overcome the voltage-time dilemma in microelectronic device applications. Further, CIPSe could potentially be driven into a quantum paraelectric phase by the application of a nominal strain.

The Department of Energy will provide public access to these results of federally sponsored research in accordance with the DOE Public Access Plan [61]. In addition, simulation inputs and outputs for the DFT calculations performed in this work are available via the Materials Data Facility [62].

## ACKNOWLEDGMENTS

This research was conducted at the Center for Nanophase Materials Sciences (CNMS), which is a DOE Office of Science User Facility and used resources of the Compute and Data Environment for Science (CADES) at ORNL. Computations also used resources of the National Energy Research Scientific Computing Center (NERSC), a U.S. Department of Energy Office of Science User Facility located at Lawrence Berkeley National Laboratory, operated under Contract No. DE-AC02-05CH11231. We thank Petro Maksymovych for discussions. This manuscript has been authored by UT-Battelle, LLC under Contract No. DE-AC05-00OR22725 with the U.S. Department of Energy. The United States Government retains and the publisher, by accepting the article for publication, acknowledges that the United States Government retains a nonexclusive, paid-up, irrevocable, worldwide license to publish or reproduce the published form of this manuscript, or allow others to do so, for United States Government purposes. P.G. and N.S. designed the project. N.S. performed all the calculations and analysis. N.S. also produced the first draft of the manuscript which was rewritten subsequently by all authors. P.G. oversaw the entire project.

[1] M. Wu and P. Jena, The rise of two-dimensional van der Waals ferroelectrics, WIREs Comput. Mol. Sci. **8**, e1365 (2018).

[2] M. Osada and T. Sasaki, The rise of 2D dielectrics/ferroelectrics, APL Mater. **7**, 120902 (2019).

[3] A. Belianinov, Q. He, A. Dziaugys, P. Maksymovych, E. Eliseev, A. Borisevich, A. Morozovska, J. Banys, Y. Vysochanskii, and S. V. Kalinin, CuInP$_2$S$_6$ room temperature layered ferroelectric, Nano Lett. 15, 3808 (2015).

[4] F. Liu, L. You, K. L. Seyler, X. Li, P. Yu, J. Lin, X. Wang, J. Zhou, H. Wang, H. He, S. T. Pantelides, W. Zhou, P. Sharma, X. Xu, P. M. Ajayan, J. Wang, and Z. Liu, Room-temperature ferroelectricity in CuInP$_2$S$_6$ ultrathin flakes, Nat. Commun. 7, 12357 (2016).

[5] J. Xiao, H. Zhu, Y. Wang, W. Feng, Y. Hu, A. Dasgupta, Y. Han, Y. Wang, D. A. Muller, L. W. Martin, P. A. Hu, and X. Zhang, Intrinsic Two-Dimensional Ferroelectricity with Dipole Locking, Phys. Rev. Lett. 120, 227601 (2018).

[6] Z. Fei, W. Zhao, T. A. Palomaki, B. Sun, M. K. Miller, Z. Zhao, J. Yan, X. Xu, and D. H. Cobden, Ferroelectric switching of a two-dimensional metal, Nature (London) 560, 336 (2018).

[7] S. Yuan, X. Luo, H. L. Chan, C. Xiao, Y. Dai, M. Xie, and J. Hao, Room-temperature ferroelectricity in MoTe$_2$ down to the atomic monolayer limit, Nat. Commun. 10, 1775 (2019).

[8] Y. Kenji, W. Xirui, W. Kenji, T. Takashi, and J.-H. Pablo, Stacking-engineered ferroelectricity in bilayer boron nitride, Science 372, eabd3230 (2021).

[9] V. Maisonneuve, V. B. Cajipe, A. Simon, R. Von Der Muhll, and J. Ravez, Ferrielectric ordering in lamellar CuInP$_2$S$_6$, Phys. Rev. B 56, 10860 (1997).

[10] W. Song, R. Fei, and L. Yang, Off-plane polarization ordering in metal chalcogen diphosphates from bulk to monolayer, Phys. Rev. B 96, 235420 (2017).

[11] A. Dziaugys, K. Kelley, J. A. Brehm, L. Tao, A. Puretzky, T. Feng, A. O'Hara, S. Neumayer, M. Chyasnavichyus, E. A. Eliseev, J. Banys, Y. Vysochanskii, F. Ye, B. C. Chakoumakos, M. A. Susner, M. A. McGuire, S. V. Kalinin, P. Ganesh, N. Balke, S. T. Pantelides *et al.*, Piezoelectric domain walls in van der Waals antiferroelectric CuInP$_2$Se$_6$, Nat. Commun. 11, 3623 (2020).

[12] Y. Qi and A. M. Rappe, Widespread Negative Longitudinal Piezoelectric Responses in Ferroelectric Crystals with Layered Structures, Phys. Rev. Lett. 126, 217601 (2021).

[13] S. M. Neumayer, E. A. Eliseev, M. A. Susner, A. Tselev, B. J. Rodriguez, J. A. Brehm, S. T. Pantelides, G. Panchapakesan, S. Jesse, S. V. Kalinin, M. A. McGuire, A. N. Morozovska, P. Maksymovych, and N. Balke, Giant negative electrostriction and dielectric tunability in a van der Waals layered ferroelectric, Phys. Rev. Mater. 3, 024401 (2019).

[14] A. Dziaugys, I. Zamaraite, J. Macutkevic, D. Jablonskas, S. Miga, J. Dec, Y. Vysochanskii, and J. Banys, Non-linear dielectric response of layered CuInP$_2$S$_6$ and Cu$_{0.9}$Ag$_{0.1}$InP$_2$S$_6$ crystals, Ferroelectrics 569, 280 (2020).

[15] J. Wu, H.-Y. Chen, N. Yang, J. Cao, X. Yan, F. Liu, Q. Sun, X. Ling, J. Guo, and H. Wang, High tunnelling electroresistance in a ferroelectric van der Waals heterojunction via giant barrier height modulation, Nat. Electron. 3, 466 (2020).

[16] M. Si, P.-Y. Liao, G. Qiu, Y. Duan, and P. D. Ye, Ferroelectric field-effect transistors based on MoS$_2$ and CuInP$_2$S$_6$ two-dimensional van der Waals heterostructure, ACS Nano 12, 6700 (2018).

[17] W. Huang, F. Wang, L. Yin, R. Cheng, Z. Wang, M. G. Sendeku, J. Wang, N. Li, Y. Yao, and J. He, Gate-coupling-enabled robust hysteresis for nonvolatile memory and programmable rectifier in van der Waals ferroelectric heterojunctions, Adv. Mater. 32, 1908040 (2020).

[18] X. Bourdon, V. Maisonneuve, V. Cajipe, C. Payen, and J. Fischer, Copper sublattice ordering in layered CuMP$_2$Se$_6$(M = In, Cr), J. Alloys Compd. 283, 122 (1999).

[19] Y. Vysochanskii, A. Molnar, M. Gurzan, V. Cajipe, and X. Bourdon, Dielectric measurement study of lamellar CuInP$_2$Se$_6$: Successive transitions towards a ferroelectric state via an incommensurate phase? Solid State Commun. 115, 13 (2000).

[20] J. Hlinka, T. Janssen, and V. Dvorák, Order-disorder versus soft mode behaviour of the ferroelectric phase transition in Sn$_2$P$_2$S$_6$, J. Phys: Condens. Matter 11, 3209 (1999).

[21] V. Liubachko, V. Shvalya, A. Oleaga, A. Salazar, A. Kohutych, A. Pogodin, and Y. M. Vysochanskii, Anisotropic thermal properties and ferroelectric phase transitions in layered CuInP$_2$S$_6$ and CuInP$_2$Se$_6$ crystals, J. Phys. Chem. Solids 111, 324 (2017).

[22] S. Aubry, A unified approach to the interpretation of displacive and order–disorder systems. I. Thermodynamical aspect, J. Chem. Phys. 62, 3217 (1975).

[23] M. A. Susner, M. Chyasnavichyus, M. A. McGuire, P. Ganesh, and P. Maksymovych, Metal thio- and selenophosphates as multifunctional van der Waals layered materials, Adv. Mater. 29, 1602852 (2017).

[24] G. Kresse and J. Furthmüller, Efficient iterative schemes for *ab initio* total-energy calculations using a plane-wave basis set, Phys. Rev. B 54, 11169 (1996).

[25] S. Grimme, Semiempirical GGA-type density functional constructed with a long-range dispersion correction, J. Comput. Chem. 27, 1787 (2006).

[26] R. D. King-Smith and D. Vanderbilt, Theory of polarization of crystalline solids, Phys. Rev. B 47, 1651 (1993).

[27] H. T. Stokes, D. M. Hatch, and B. J. Campbell, Isotropy software suite, iso.byu.edu.

[28] R. Pfeiff and R. Kniep, Quaternary selenodiphosphates(iv): M$^I$M$^{III}$[P$_2$Se$_6$], (M$^I$ = Cu, Ag; M$^{III}$ = Cr, Al, Ga, In), J. Alloys Compd. 186, 111 (1992).

[29] A disorder phase with the space group $P\bar{3}1c$ has also been proposed as the high-symmetry PE phase [18]. As the primary purpose of this letter is to understand the microscopic origin and the stabilization of local dipoles in CIPSe, we use the ordered high-symmetry phase as the parent structure.

[30] X. Li, T. Cao, Q. Niu, J. Shi, and J. Feng, Coupling the valley degree of freedom to antiferromagnetic order, Proc. Natl. Acad. Sci. U.S.A. 110, 3738 (2013).

[31] N. Sivadas, M. W. Daniels, R. H. Swendsen, S. Okamoto, and D. Xiao, Magnetic ground state of semiconducting transition-metal trichalcogenide monolayers, Phys. Rev. B 91, 235425 (2015).

[32] H.-G. Unruh, The spontaneous polarization of (NH$_4$)$_2$SO$_4$, Solid State Commun. 8, 1951 (1970).

[33] While one might conclude from Fig. 3 that the energy-surface for In displacement (orange) is flatter compared to the Cu displacement (blue), this is not the case. The energy reported is with respect to the fractional shift of the atoms. From Table III, we can see that the amplitude of the shift of the Cu atoms is more than six times that of the In atoms.

[34] See Supplemental Material at http://link.aps.org/supplemental/10.1103/PhysRevResearch.4.013094 for technical details of the calculation and additional information.

[35] R. E. Cohen, Origin of ferroelectricity in perovskite oxides, Nature (London) **358**, 136 (1992).

[36] K. Z. Rushchanskii, Y. M. Vysochanskii, and D. Strauch, Ferroelectricity, Nonlinear Dynamics, and Relaxation Effects in Monoclinic $Sn_2P_2S_6$, Phys. Rev. Lett. **99**, 207601 (2007).

[37] J. Holakovský, A new type of the ferroelectric phase transition, Phys. Status Solidi B **56**, 615 (1973).

[38] C. J. Fennie and K. M. Rabe, Ferroelectric transition in $YMnO_3$ from first principles, Phys. Rev. B **72**, 100103(R) (2005).

[39] V. A. Isupov, Ferroelectric and ferroelastic phase transitions in molybdates and tungstates of monovalent and bivalent elements, Ferroelectrics **322**, 83 (2005).

[40] N. A. Benedek and C. J. Fennie, Hybrid Improper Ferroelectricity: A Mechanism for Controllable Polarization-Magnetization Coupling, Phys. Rev. Lett. **106**, 107204 (2011).

[41] H. J. Zhao, M. N. Grisolia, Y. Yang, J. Íñiguez, M. Bibes, X. M. Chen, and L. Bellaiche, Magnetoelectric effects via pentalinear interactions, Phys. Rev. B **92**, 235133 (2015).

[42] J. Íñiguez, M. Stengel, S. Prosandeev, and L. Bellaiche, First-principles study of the multimode antiferroelectric transition in $PbZrO_3$, Phys. Rev. B **90**, 220103(R) (2014).

[43] B. Xu, O. Hellman, and L. Bellaiche, Order-disorder transition in the prototypical antiferroelectric $PbZrO_3$, Phys. Rev. B **100**, 020102(R) (2019).

[44] X. Lu, H. Li, and W. Cao, Landau expansion parameters for $BaTiO_3$, J. Appl. Phys. **114**, 224106 (2013).

[45] K. M. Rabe, C. H. Ahn, and J. Triscone, *Physics of Ferroelectrics: A Modern Perspective*, Vol. 105 (Springer Science and Business Media, Berlin, 2007).

[46] S. Lee, P. Colombet, G. Ouvrard, and R. Brec, General trends observed in the substituted thiophosphate family. synthesis and structure of silver scandium thiophosphate, $AgScP_2S_6$, and cadmium iron thiophosphate, $CdFeP_2S_6$, Inorg. Chem. **27**, 1291 (1988).

[47] S.-H. Wei, S. B. Zhang, and A. Zunger, Off-Center Atomic Displacements in Zinc-Blende Semiconductor, Phys. Rev. Lett. **70**, 1639 (1993).

[48] J. K. Burdett and O. Eisenstein, From three- to four-coordination in copper(i) and silver(i), Inorg. Chem. **31**, 1758 (1992).

[49] J. H. Haeni, P. Irvin, W. Chang, R. Uecker, P. Reiche, Y. L. Li, S. Choudhury, W. Tian, M. E. Hawley, B. Craigo, A. K. Tagantsev, X. Q. Pan, S. K. Streiffer, L. Q. Chen, S. W. Kirchoefer, J. Levy, and D. G. Schlom, Room-temperature ferroelectricity in strained $SrTiO_3$, Nature (London) **430**, 758 (2004).

[50] K. A. Müller and H. Burkard, $SrTiO_3$: An intrinsic quantum paraelectric below 4 k, Phys. Rev. B **19**, 3593 (1979).

[51] U. Aschauer and N. A. Spaldin, Competition and cooperation between antiferrodistortive and ferroelectric instabilities in the model perovskite $SrTiO_3$, J. Phys.: Condens. Matter **26**, 122203 (2014).

[52] I. Zamaraite, V. Liubachko, R. Yevych, A. Oleaga, A. Salazar, A. Dziaugys, J. Banys, and Y. Vysochanskii, Quantum paraelectric state and critical behavior in $Sn(Pb)_2P_2S(Se)_6$ ferroelectrics, J. Appl. Phys. **128**, 234105 (2020).

[53] R. Merlin, Generating coherent THz phonons with light pulses, Solid State Commun. **102**, 207 (1997).

[54] R. I. Tobey, D. Prabhakaran, A. T. Boothroyd, and A. Cavalleri, Ultrafast Electronic Phase Transition in $La_{1/2}Sr_{3/2}MnO_4$ by Coherent Vibrational Excitation: Evidence for Nonthermal Melting of Orbital Order, Phys. Rev. Lett. **101**, 197404 (2008).

[55] M. Först, C. Manzoni, S. Kaiser, Y. Tomioka, Y. Tokura, R. Merlin, and A. Cavalleri, Nonlinear phononics as an ultrafast route to lattice control, Nat. Phys. **7**, 854 (2011).

[56] T. Schenk, M. Pešić, S. Slesazeck, U. Schroeder, and T. Mikolajick, Memory technology—A primer for material scientists, Rep. Prog. Phys. **83**, 086501 (2020).

[57] S. Li and T. Birol, Suppressing the ferroelectric switching barrier in hybrid improper ferroelectrics, npj Comput. Mater. **6**, 168 (2020).

[58] A. C. Garcia-Castro, N. A. Spaldin, A. H. Romero, and E. Bousquet, Geometric ferroelectricity in fluoroperovskites, Phys. Rev. B **89**, 104107 (2014).

[59] K. F. Garrity, K. M. Rabe, and D. Vanderbilt, Hyperferroelectrics: Proper Ferroelectrics with Persistent Polarization, Phys. Rev. Lett. **112**, 127601 (2014).

[60] A. C. Garcia-Castro, W. Ibarra-Hernandez, E. Bousquet, and A. H. Romero, Direct Magnetization-Polarization Coupling in $BaCuF_4$, Phys. Rev. Lett. **121**, 117601 (2018).

[61] http://energy.gov/downloads/doe-public-access-plan.

[62] N. Sivadas, P. Doak, and P. Ganesh, Dataset for anharmonic stabilization of ferrielectricity in $CuInP_2Se_6$, Mater. Data Facility (2021).